# Measurement of the CMS Magnetic Field

V. I. Klyukhin, *Member, IEEE*, A. Ball, F. Bergsma, D. Campi, B. Curé, A. Gaddi, H. Gerwig, A. Hervé, J. Korienek, F. Linde, C. Lindenmeyer, R. Loveless, M. Mulders, T. Nebel, R. P. Smith, D. Stickland, G. Teafoe, L. Veillet, and J. K. Zimmerman

*Abstract*—The measurement of the magnetic field in the tracking volume inside the superconducting coil of the Compact Muon Solenoid (CMS) detector under construction at CERN is done with a fieldmapper designed and produced at Fermilab. The fieldmapper uses 10 3-D B-sensors (Hall probes) developed at NIKHEF and calibrated at CERN to precision 0.05% for a nominal 4 T field. The precise fieldmapper measurements are done in 33840 points inside a cylinder of 1.724 m radius and 7 m long at central fields of 2, 3, 3.5, 3.8, and 4 T. Three components of the magnetic flux density at the CMS coil maximum excitation and the remanent fields on the steel-air interface after discharge of the coil are measured in check-points with 95 3-D B-sensors located near the magnetic flux return yoke elements. Voltages induced in 22 flux-loops made of 405-turn installed on selected segments of the yoke are sampled online during the entire fast discharge (190 s time-constant) of the CMS coil and integrated offline to provide a measurement of the initial magnetic flux density in steel at the maximum field to an accuracy of a few percent. The results of the measurements made at 4 T are reported and compared with a three-dimensional model of the CMS magnet system calculated with TOSCA.

## I. INTRODUCTION

THE Compact Muon Solenoid (CMS) is a general-purpose detector designed to run at the highest luminosity at the CERN Large Hadron Collider (LHC). Its distinctive features include a 4 T superconducting solenoid with 6 m diameter by 12.5 m long free bore, enclosed inside a 10 000-ton yoke made of construction steel: five dodecagonal three-layered barrel wheels and three end-cap disks at each end, comprised of steel plates up to 620 mm thick, which return the flux of the solenoid and serve as the absorber plates of the muon detection system [1], [2]. The yoke steel contains up to 0.17% C, up to 1.22% Mn, and also some Si, Cr, and Cu.

A three-dimensional (3-D) model of the magnetic field of the CMS magnet has been prepared [3] for utilization during the engineering phase of the magnet system and early physics studies of the anticipated performance of the detector, as well as for track parameter reconstruction when the detector begins operation.

To reduce the uncertainty in utilization of the calculated values for the magnetic field, which is used to determine the momenta of muons during detector operation, a direct measurement of the average magnetic flux density in selected regions of the yoke by an integration technique is done with 22 flux-loops made of 405-turn installed around selected CMS yoke plates.

The areas enclosed by the flux-loops vary from 0.3 to 1.58 $m^2$ on the barrel wheels and from 0.5 to 1.13 $m^2$ on the end-cap disks. The flux-loops measure the variations of the magnetic flux induced in the steel when the field in the solenoid is changed during the "fast" (190 s time constant) discharge made possible by the protection system provided to protect the magnet in the event of major faults [4], [5]. The test of the protection system during the commissioning of the CMS magnet provided the opportunity for the flux-loop measurements. The system of 80 3-D Hall probes (out of 95 probes installed) measured the remanent field on the disk steel-air interface to be added to the flux-loop measurements after full discharge of the CMS coil.

To investigate if the measurements of the average magnetic flux density in the CMS yoke plates could be done with accuracy of a few percent using flux-loops, a special R&D program was performed with sample disks made of the CMS yoke steel from different melts [6], [7].

Those studies showed that the contribution of eddy currents to the voltages induced in the test flux-loop is negligible. The contribution of eddy currents to the voltages induced in the flux-loops installed on the CMS magnet yoke when the "fast" discharge of the CMS coil occurs was investigated also [8] with Vector Fields' program ELEKTRA [9].

The precise measurement of the magnetic field in the tracking volume inside the CMS coil is done with a fieldmapper designed and produced at Fermilab. The fieldmapper uses 10 3-D B-sensors (Hall probes) developed at NIKHEF and calibrated at CERN to precision $5 \cdot 10^{-4}$ at 4.5 T field.

Monitoring the CMS magnetic field is done with 4 NMR probes (Model 1062-R) of the METROLAB PT 2025 high precision Teslameter [10]. The probes are installed near the inner wall of the superconducting coil cryostat. Two probes of the same type were also used in the fieldmapping to measure the field along the coil axis, and at the largest radius of the measured volume.

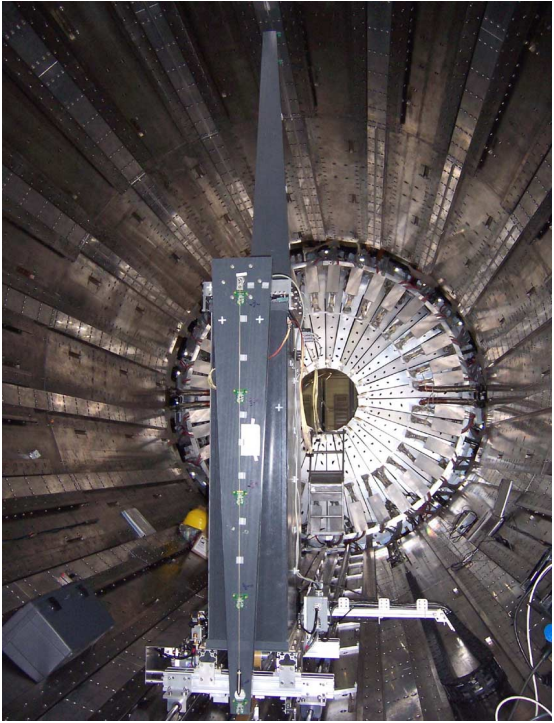

Fig. 1. The CMS fieldmapper mounted inside the hadronic barrel calorimeter inner volume.

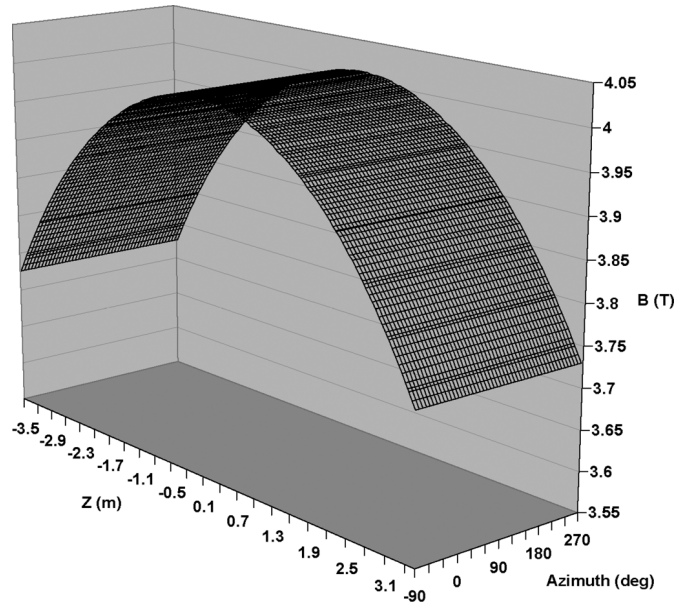

Fig. 2. Magnetic flux density measured at radius 92 mm along the coil axis in the range of $\pm 3.5$ m with respect to the coil middle plane for full azimuth coverage.

## II. THE FIELDMAPPER DESCRIPTION

The volume mapped by the fieldmapper inside the CMS coil is a cylinder of 1.724 m radius and 7 m long. The fieldmapper shown in Fig. 1 inside the measured volume moves along the rails installed along the coil axis in the hadronic barrel calorimeter, stopping at predefined points where two sensor arms can be rotated through 360°, stopping at predefined angles where the magnetic field is sampled. The azimuth steps are 7.5° in magnitude. Steps along the coil axis are fixed to 50 mm by a tensioned toothed Kevlar belt.

Each arm of the fieldmapper contains 5 3-D B-sensors located at radii 92, 500, 908, 1316, and 1724 mm off the coil axis. The distance between the negative and positive arm B-sensors along the coil axis that coincides with the CMS Z-axis, is 950 mm. Therefore, making 19 steps along the coil axis in positive direction, the fieldmapper delivers the B-sensors of negative arm in the same Z-position where B-sensors of positive arm were before and vice versa.

Made of nonmagnetic materials, the fieldmapper uses pneumatic power, the gas flow is controlled with 24-V piezoelectric valves, the remote operation is performed via a Programmable Logic Controller and operator's LabVIEW [11] console, and the laser ranger is used for absolute Z-coordinate reference after unscheduled stop.

The alignment of the fieldmapper azimuth axle with respect to the CMS coil axis is performed with a precision better than 1.9 mrad. The read-out of the B-sensors is performed via the CANopen protocol [12], [13].

## III. MEASURED AND MODELED CMS MAGNETIC FIELD COMPARISON

### A. Mapping the Inner Coil Volume

Fieldmapping the inner coil volume is done at five different values of the magnetic flux density in the center of the coil: 2, 3, 3.5, 3.8, and 4 T. The dependence of the central magnetic flux density $B_0$ measured with the NMR probes on the coil current is linear in the range of the coil current from 4 to 19.14 kA that corresponds to the range of $B_0$ from 0.85 to 4 T.

At each of these $B_0$ values the magnetic flux density is mapped in 141 azimuth planes, at 48 azimuth angles, thus the full number of the points mapped with 5 B-sensors is 33840. The central part of the volume in Z-range of $\pm 2.55$ m with respect to the coil middle plane is mapped twice in the same pass of the fieldmapper through the volume with B-sensors of positive and negative arms. The difference between magnetic flux density B measured in the same point with the B-sensor of positive and negative arm does not exceed 1 mT.

In Fig. 2 the magnetic flux density B measured at 4 T central field near the coil axis with B-sensors located at radii 92 mm is displayed without any corrections for the B-sensor misalignment. This plot demonstrates the high quality of the measurements and shows no variations of B with the azimuth angle. The general precision of measurements is 0.07%.

The measurements discovered a very small asymmetry of the magnetic flux density with respect to the coil middle plane caused by one missing turn out of 2180 designed turns.

### B. Measured and Calculated Magnetic Field Comparison

To explain the observed magnetic field asymmetry a full CMS length 3-D model is calculated with Vector Fields' program TOSCA [14]. The model is based on a half-volume of the CMS

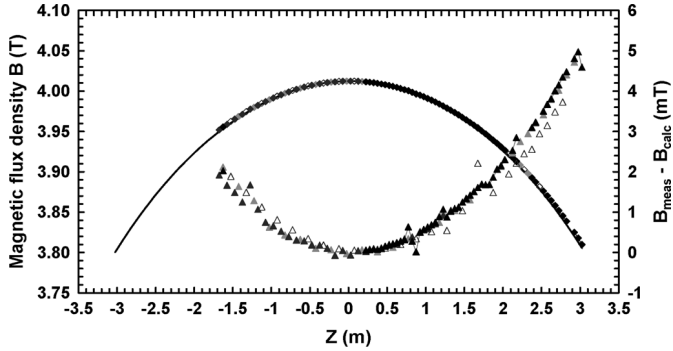

Fig. 3. Magnetic flux density (left scale) measured at the coil axis with NMR probe (rhombs) in the range from $-1.675$ to $3.025$ m with respect to the coil middle plane. Smooth curve represents the calculations done with the CMS TOSCA model. The triangles show the difference between the measured and calculated B values (right scale). The model is normalized to get the measured $B_0$ value 4.0124 T. Different colors mark four different sets of the measurements when the fieldmapper traveled in both axial directions.

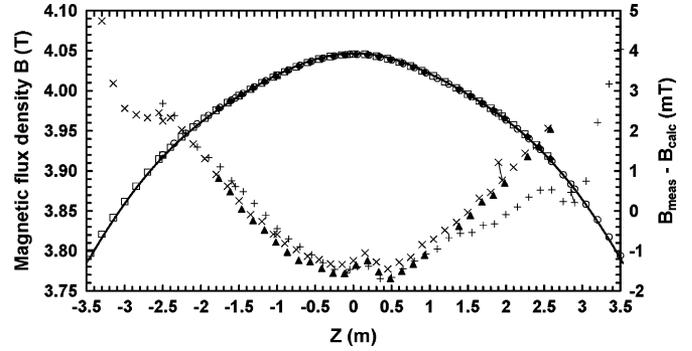

Fig. 4. Magnetic flux density (left scale) measured along the coil axis in horizontal plane at radius 1.724 m in the range of $\pm 3.5$ m with respect to the coil central plane with the NMR probe, and two B-sensors located on negative (open squares) and positive (open circles) fieldmapper arms, respectively. Smooth curve represents the calculations done with the CMS TOSCA model. Triangles show the difference (right scale) between the NMR measurements and calculations (the range of measurements is from $-1.767$ to $2.583$ m), slanted crosses present the difference between the negative arm B-sensor measurements and calculations, and right crosses display the difference between the positive arm B-sensor measurements and calculations. The model is normalized to get the measured $B_0$ value 4.0124 T.

steel yoke and contains 21 coil conductors to model the missing turn and 1922958 nodes.

For each set of the measurements the current in the model is normalized to get the $B_0$ value measured with NMR probe or averaged B value measured with B-sensors in the coil middle plane at radius 92 mm. Then the measured and calculated B values are compared along Z-direction.

The NMR probe measurements are done on the axis at two $B_0$ values: 3 and 4 T. The NMR probe measurement is also done in horizontal plane at maximum radius 1.724 m at the central field of 4 T. The field gradient and noise conditions allowed measuring B values on the axis in Z-range from $-1.675$ to $3.025$ m and measuring B values at the maximum radius in Z-range from $-1.767$ to $2.583$ m with NMR-probes. At the maximum radius both negative and positive arm B-sensors covered the full Z-range of $\pm 3.5$ m.

In Fig. 3 the comparison between the NMR measurements on the coil axis and the calculated values is displayed for the central field 4.0124 T. The triangles of different color show the difference between the measured and calculated B values: the maximum difference reaches 5 mT when the measured field is 3.8178 T. Four sets of the NMR measurements do not differ more than by 0.5 mT.

In Fig. 4 the comparison between the NMR and B-sensors measurements done at the maximum radius and the calculated values is also displayed for the central field 4.0124 T. The black triangles show the difference between the NMR measurements and calculated B values: the maximum difference reaches 2 mT when the measured field is 3.9152 T. The slanted and right crosses show the difference between the B-sensors measurements and calculated B values: the maximum difference reaches 5 mT when the measured field is 3.7974 T. Both NMR and B-sensors measurements consist well. Thus, the CMS 3-D model with one missing turn explains well the very small magnetic field asymmetry observed in the measurements. No axial shift of the coil with respect to the yoke is required to fit the measurements and the calculations.

This conclusion is confirmed by the coil alignment and the coil cryostat position measurements.

## IV. THE FLUX-LOOP MEASUREMENTS

Measuring the averaged magnetic flux density inside the steel blocks of the CMS yoke is performed with the system of 22 flux-loops made of 405-turns mounted around the selected elements of the yoke. During the coil "fast" discharge (190 s time-constant) from the full current to zero, voltages with amplitudes up to 3–4.5 V were induced in the loops.

The digitization of the flux-loop voltages is done with seven Measurement Computing USB-based DAQ modules USB-1208LS with 4 differential 12-bit analog inputs each [15].

The off-line integration of the voltages and adding the remanent fields reconstruct the maximum magnetic flux density in steel blocks at the CMS magnet full excitation. The values of the remanent fields are below 10 mT.

In Fig. 5 the induced voltages and reconstructed magnetic flux densities are shown for three flux-loops mounted in three iron layers between the muon drift chambers in the lower azimuth sector of the next to the central barrel wheel of the yoke. The areas enclosed by the flux-loops are 0.38 m$^2$ (inner layer), 1.29 m$^2$ (middle layer), and 1.58 m$^2$ (outer layer). The minimum area enclosed by the flux-loop on the barrel wheels of the yoke is 0.3 m$^2$, the areas enclosed by the flux-loops on the end-cap disks of the yoke vary from 0.5 to 1.13 m$^2$.

At the central field 4.0124 T the magnetic flux densities reconstructed with the flux-loops in the barrel wheels of the yoke vary from 0.62 to 1.97 T. The magnetic flux density measured with the flux-loops in the end-cap disks of the yoke vary from 1.66 to 2.62 T. The precision of measurements is expected to be a few per cent.

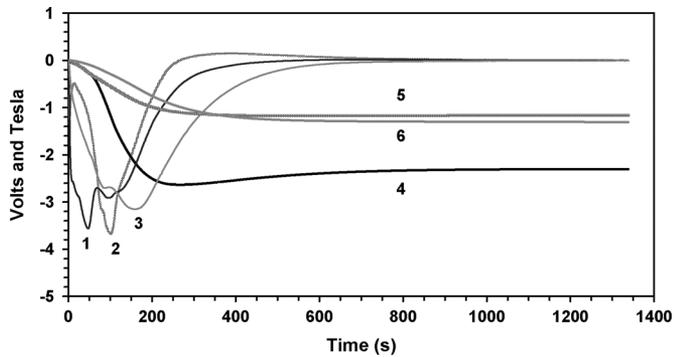

Fig. 5. Voltages (curves 1, 2, and 3) induced in steel blocks of the barrel next to central wheel in the inner (1), middle (2), and outer (3) layers during the coil "fast" discharge from full current of 19140 A to zero and integrated magnetic flux densities (curves 4, 5, and 6) in the inner (4), middle (5), and outer (6) layers.

## V. CONCLUSION

The first measurements of the magnetic flux density in the volume of more than 65 $m^3$ inside the superconducting solenoid coil at 4 T central field is done with precision of 0.07%.

The developed 3-D model of the CMS magnet describes well the magnetic flux distribution over all the magnet volume and is used outside the CMS coil to provide the magnetic field for the simulation and tracking reconstruction programs.

The developed flux-loops technique permits to measure the average magnetic flux densities in steel elements of the yoke with precision of a few per cent.

The response of the CMS coil central field to the coil current is linear within a current range from 4 to 19.14 kA.

Monitoring the magnetic flux density in different regions of the magnet yoke with 3-D B-sensors developed at NIKHEF is stable and provides the magnetic field values with a high precision. Monitoring the magnetic flux density inside the coil with NMR probes of three different types gives the reference magnetic field values in all the range required by the CMS detector performance for the particle physics.